\documentclass[tightenlines,aps,prc,showpacs,amsmath,showkeys,
superscriptaddress]{revtex4}
\usepackage[dvips]{graphicx}
\usepackage{overpic}

\begin{document}

\title{\bf Unified approach to nuclear densities
from exotic atoms}

\author{E.~Friedman}
\email{elifried@vms.huji.ac.il}
\affiliation{Racah Institute of Physics, The Hebrew University,
Jerusalem 91904, Israel}

\date{\today}

\begin{abstract}
Parameters of nuclear density distributions are derived from 
least-squares fits to strong
interaction  observables in exotic atoms. Global analyses of
antiprotonic and pionic atoms show reasonably good agreement between
the two types of probes regarding the average behaviour of root-mean-square
radii of the neutron distributions. Apparent conflict regarding the shape of
the neutron distribution is attributed to different radial
sensitivities of these two probes.  
\end{abstract}

\maketitle

\section{Introduction}
\label{intro}
The density distribution of protons in nuclei is considered known as
it is obtained from the nuclear charge distribution   %~\cite{FBH95} 
  by unfolding the finite size of the charge of the proton. The neutron
distributions are, however, generally not known to sufficient accuracy.
A host of different methods has been applied in studies of
root-mean-square (rms) radii of neutron distributions in nuclei but the
results are sometimes conflicting, see e.g. \cite{BFG89,JTL04}.
In the present work we focus on antiprotonic and on pionic atoms as 
a source of information on neutron densities. We deduce average properties
with the help of global analyses of about 100 data points in each case,
covering the whole of the periodic table.  Reasonably good
agreement is obtained between the two types of exotic atoms when considering
 rms radii of the neutron distributions. A conflict
regarding the shape of the distributions is most likely due to the 
different radial sensitivities of the two probes.

\section{Method}
\label{sec:1}

Strong interaction level shifts and widths in exotic atoms, formed
by the capture of a negatively charged hadron into an atomic
orbit, are calculated with the help of an optical potential
inserted into the appropriate wave equation.
The simplest class of optical potentials
$V_{\rm opt}$ is the generic $t\rho(r)$ potential, which for underlying
$s$-wave hadron-nucleon interactions assumes the form:

\begin{equation}\label{eq:Vopt}
2\mu V_{\rm opt}(r) = - 4\pi(1+\frac{A-1}{A}\frac{\mu}{M})
\{b_0[\rho_n(r)+\rho_p(r)] + \tau_z b_1[\rho_n(r)-\rho_p(r)] \} \;.
\end{equation}
Here, $\rho_n$ and $\rho_p$ are the neutron and proton density
distributions normalized to the number of neutrons $N$ and number of protons
$Z$, respectively, $M$ is the mass of the nucleon and $\tau_z = +1$ for the
negatively charged hadrons considered in the present work.
When handling many different nuclei over the periodic table it is 
necessary to represent the densities by approximate
distributions, usually chosen as the the two-parameter Fermi 
distribution (2pF).
With the proton densities considered known, we focuse on the differences
between the neutron and the proton distributions. 
A linear dependence of $r_n-r_p$, the difference between the rms radii,
on $(N-Z)/A$ has been employed in $\bar p$ studies \cite{TJL01,JTL04,FGM05},
namely
\begin{equation} \label{eq:RMF}
r_n-r_p = \gamma \frac{N-Z}{A} + \delta \; ,
\end{equation}
with $\gamma$ close to 1.0~fm and $\delta$ close to zero. This 
parameterization is adopted here. In order to allow for possible differences in the shape of the neutron
distribution, the `skin' and `halo' forms of Ref.~\cite{TJL01} were
used, as well as an average between the two. We adopt a 2pF
distribution both for the proton (unfolded from the charge distribution)
and for the neutron density  distributions
\begin{equation}
\label{eq:2pF}
\rho_{n,p}(r)  = \frac{\rho_{0n,0p}}{1+{\rm exp}((r-R_{n,p})/a_{n,p})} \; .
\end{equation}
Then for each value of $r_n-r_p$ in the `skin' form the same diffuseness
parameter for protons and neutrons, $a_n=a_p$, is used and the
$R_n$ parameter is determined from the rms radius $r_n$. In the `halo'
form the same radius parameter, $R_n=R_p$, is assumed and $a_n^{\rm h}$
is determined from $r_n$. In the `average' option the diffuseness parameter
is set to be the average of the above two diffuseness parameters,
$a_n^{{\rm ave}}=(a_p+a_n^{\rm h})/2$, and the radius parameter $R_n$ is
then determined from the rms radius $r_n$. In this way we can test three
shapes of the neutron distribution for each value of its rms radius all
along the periodic table. These shapes provide sufficient difference
in order to be tested in global fits. The results below are presented as
the best fit  $\chi ^2$  values {\it vs.} the neutron rms radius 
parameter $\gamma$.

\section{Results}
\label{sec:3}
\subsection{Antiprotonic atoms}
\label{sec:3a}

\begin{figure}[t]
\centering
\begin{tabular}{cp{1cm}c}
\begin{overpic}[height=8cm,width=0.40\textwidth,clip]{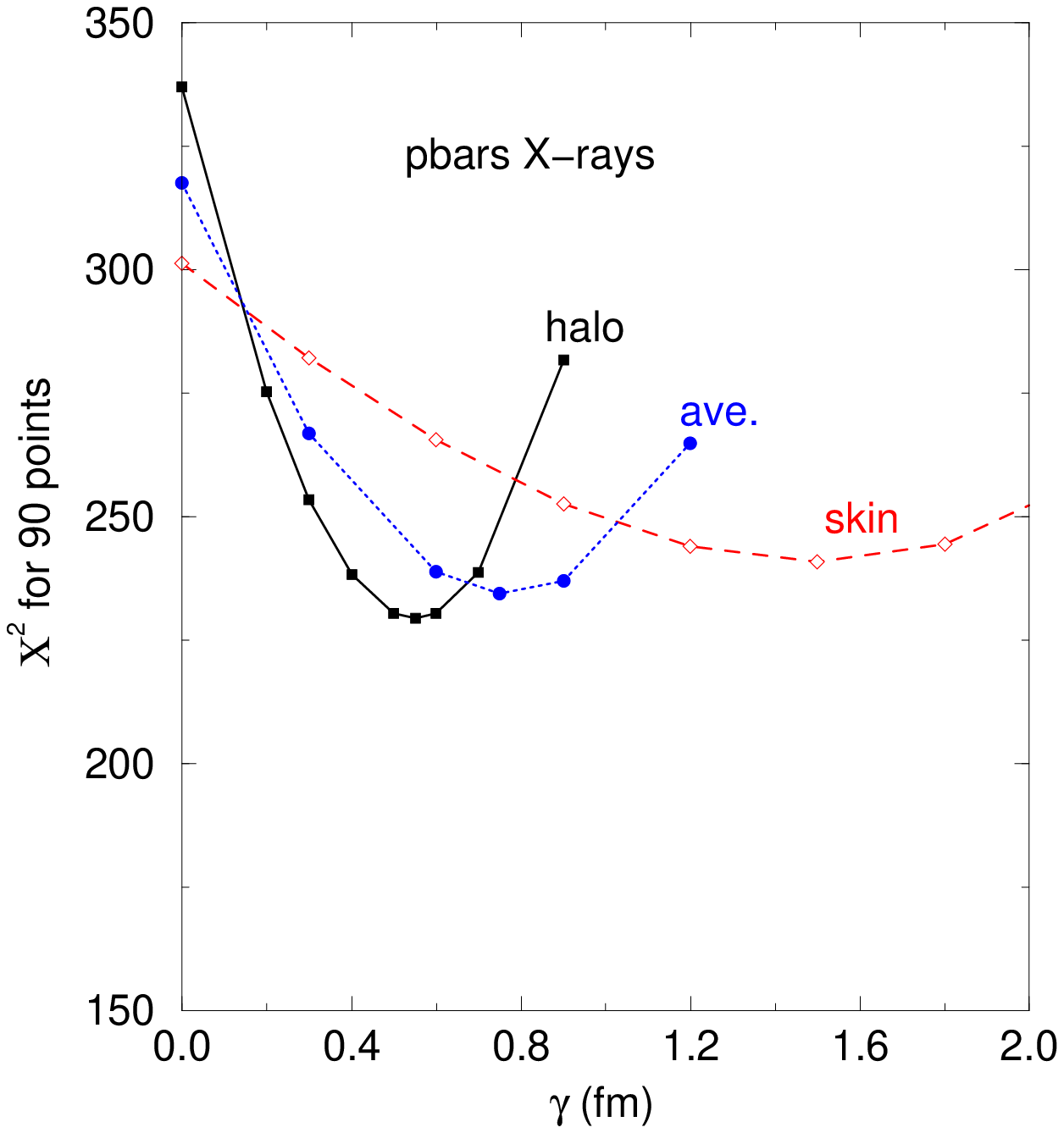}
\end{overpic}
%& &
\begin{overpic}[height=8cm,width=0.40\textwidth,clip]{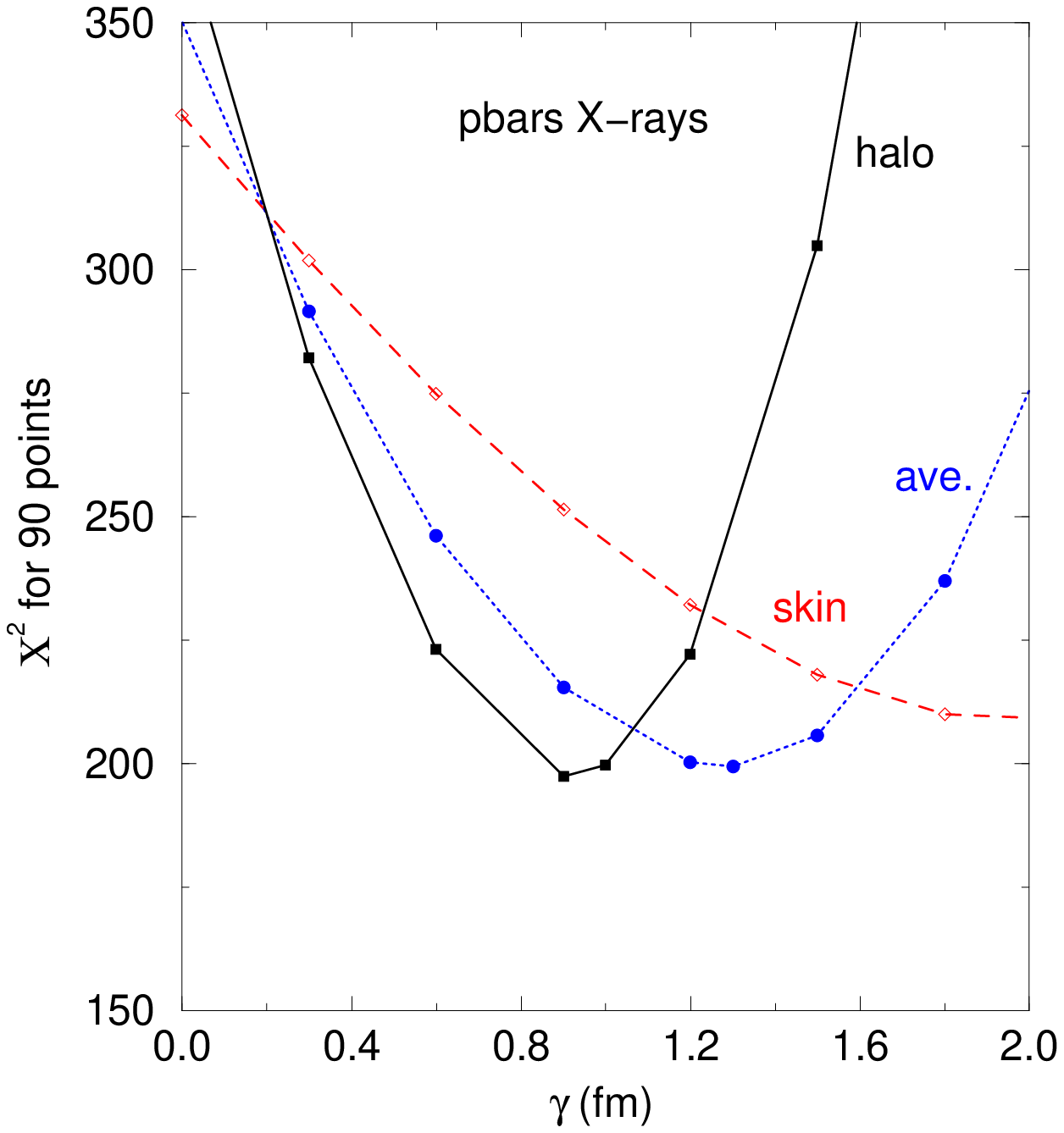}
\end{overpic}
%\\[3.5ex]
%(a) & & (b)
\end{tabular}
\caption{Zero range (left) and finite range (right) fits to strong interaction
shifts and widths in antiprotonic atoms. For the latter the best fit is
obtained for a rms radius of the $\bar p$N interaction of 1.1$\pm$0.1 fm.}
\label{fig:1}
\end{figure}

Figure \ref{fig:1} shows results of global fits to 90 data points
from measurements of X-rays by the PS209 collaboration \cite{TJC01}.
It is seen that the `skin' shape for the neutron density distribution
is unfavoured and that a finite range for the $\bar p$N interaction
leads to significant improvements in the fits. 
Comparing the zero-range (ZR) results of the left-hand side of Fig.\ref{fig:1}
with the corresponding ZR results of Ref.\cite{JTL04} we note that the values
of the best-fit parameter $\gamma$ are quite different. This is due to the
use in Ref.\cite{JTL04} of {\it fixed} values for the complex parameter $b_0$ 
for all values of $\gamma$
whereas we re-fit these phenomenological parameters when $\gamma$ changes.
In fact, with the fixed values for $b_0$ 
taken from Ref.\cite{BFG95} one obtains a value for 
$\gamma$ that merely represents an average over the neutron densities used
in Ref.\cite{BFG95} to derive those fixed values.   

The best fit finite-range potential produces 
a $\chi ^2$ per degree of freedom of about 2, which is most acceptable
considering the simplicity of the model and the extent of the data.
We also note that on the basis of values of $\chi ^2$ it is impossible
to distinguish between the `halo' and the `average' shapes, which lead to
somewhat different values of the rms radius parameter $\gamma$.
Similar analyses of the radio-chemical data and of the combined X-rays
and radiochemical data lead to very similar conclusions. 
It is interesting to note that the isovector parameter $b_1$ turns out
to be consistent with zero \cite{FGM05}.
This is in full agreement with the result of analysing separately the
radio-chemical data where we find that the best fit is obtained for 
a ratio of 0.99$\pm$0.07 for the absorption on a neutron to the absorption
on a proton.
\subsection{Pionic atoms}
\label{sec:3b}
Values of rms radii of neutron distributions for several
nuclides had been derived from pionic atoms more than two decades
ago \cite{BFG89}. The first to derive rms radii from
 extensive data sets of strong interaction
observables in pionic atoms were Garc\'{\i}a-Recio et al. \cite{GNO92}.
Using  the `skin' shape for the neutron densities they 
presented a list of rms radii which, when interpreted with the present
formulation, leads to $\gamma=1.06\pm0.34$ fm for their semi-theoretical
model and $\gamma=0.97\pm0.12$ fm for the phenomenological model of Meirav
et al. \cite{MFJ89}.

\begin{figure}[t]
\centering
\begin{tabular}{cp{1cm}c}
\begin{overpic}[height=8cm,width=0.47\textwidth,clip]{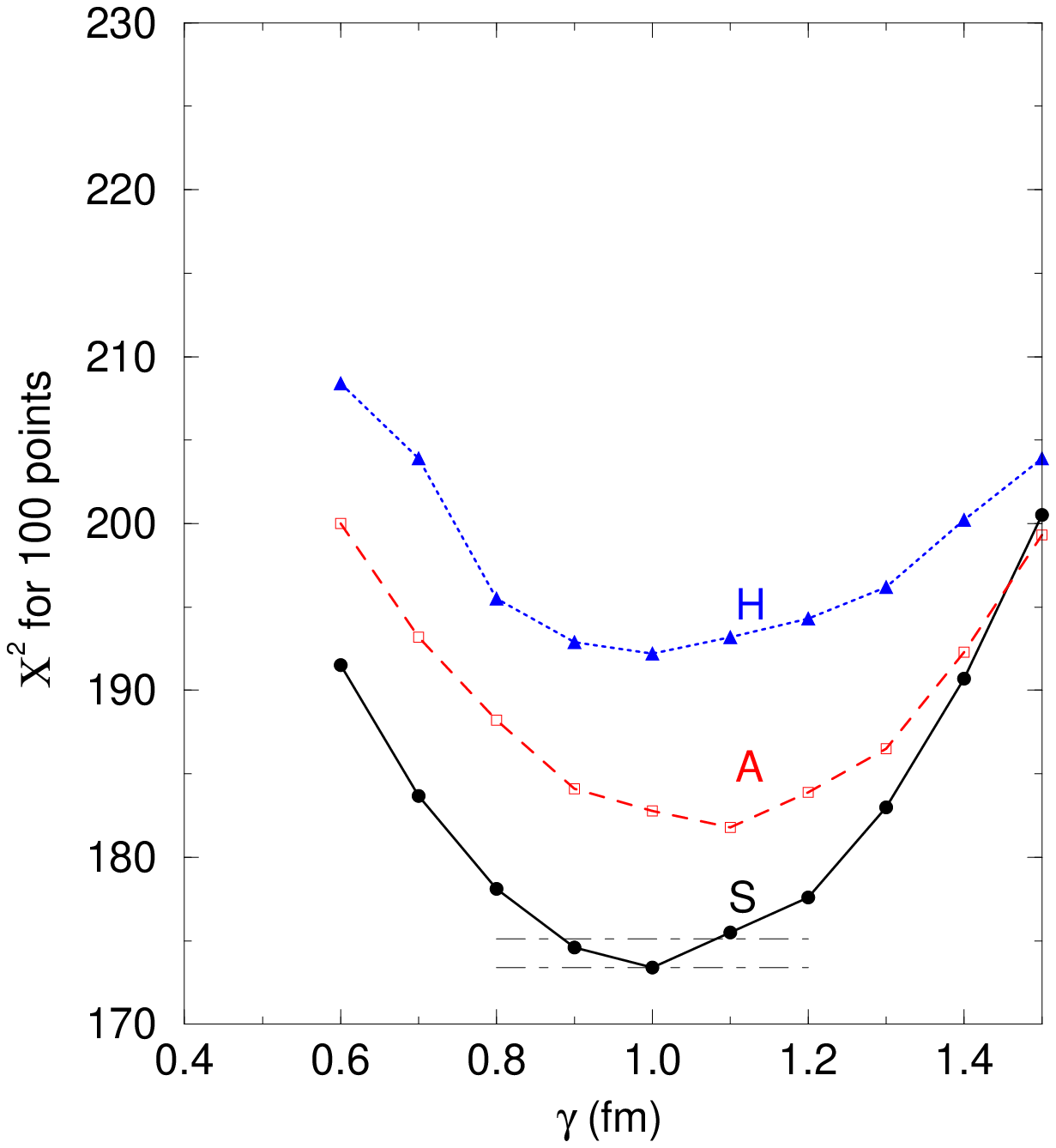}
\end{overpic}
%& &
\begin{overpic}[height=8cm,width=0.47\textwidth,clip]{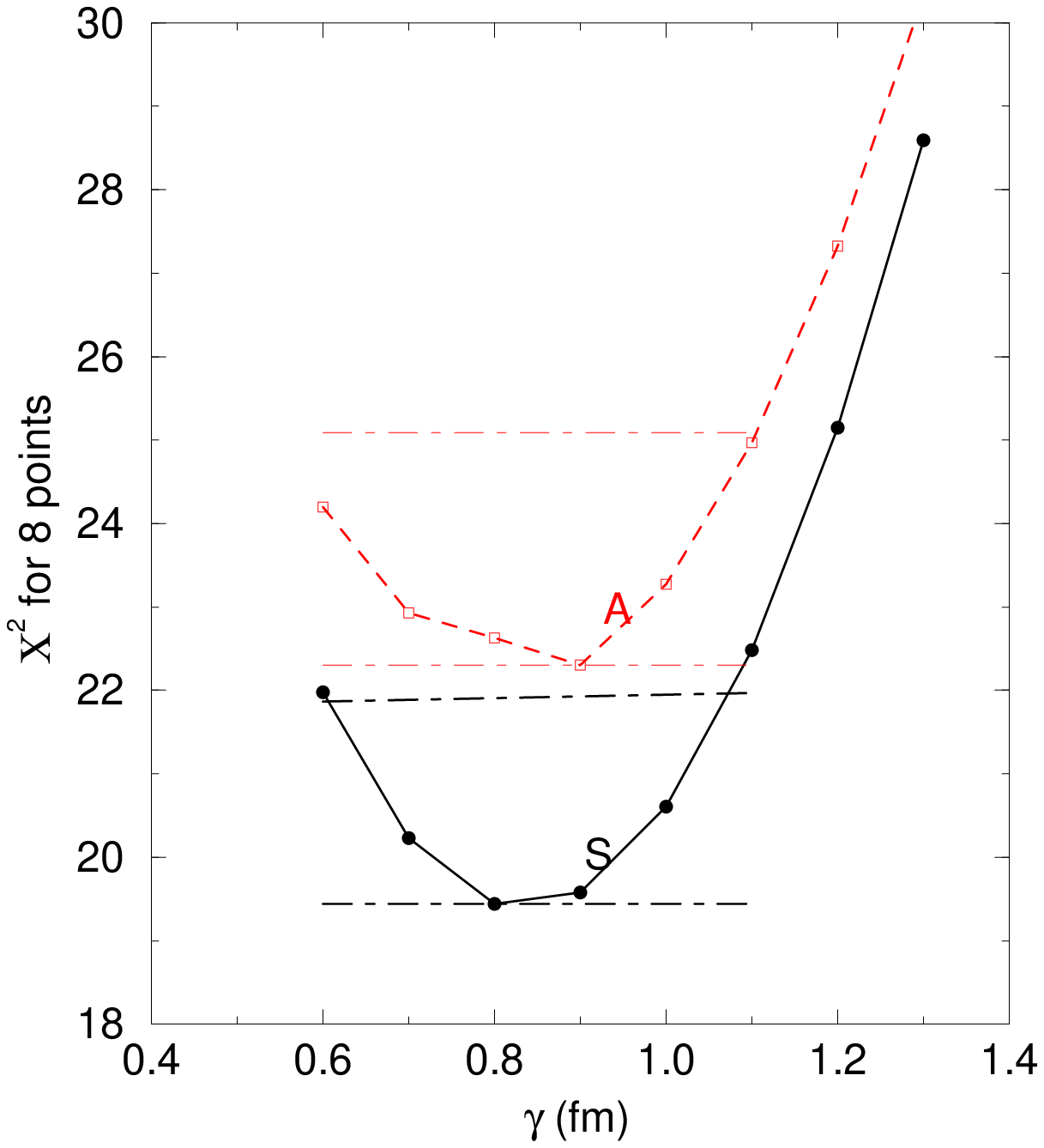}
\end{overpic}
%\\[3.5ex]
%(a) & & (b)
\end{tabular}
\caption{Left: global fits to pionic atom data. 
S, A and H represent the skin, average and halo 
shapes, respectively. The horizontal band
represents $\chi ^2$ per point, see text. Right: similar to the left
but for pionic atoms of Pb.}

\label{fig:2}
\end{figure}

Figure \ref{fig:2} shows results of global fits to 100 data points
for pionic atoms, using the latest pion-nucleus potential with 
energy and density dependence in the $s$-wave term and finite range 
in the $p$-wave term, see \cite{FGa07} for details.
The left hand side shows that the `skin' shape for the neutron density
distributions yields the lowest $\chi ^2$ value. 
The horizontal band represents the
value of $\chi ^2$ per degree of freedom, which indicates the statistical
significance of the fits and determines the uncertainty of the derived 
parameters, $\gamma$ in this case. On the right hand side we compare
data for Pb with predictions made with the global parameters of the left
hand side. The results are consistent with the global analyses but the
uncertainties are obviously considerably larger. This is typical also
of antiprotonic atoms, demonstrating the limited accuracy of a single
element analysis.

\section{Discussion}
\label{sec:4}

Table \ref{tab:1} summarizes the above results regarding the shapes of
the neutron density distributions, within the simple 2pF parameterization,
and the values of the parameter $\gamma$ of Eq.(\ref{eq:RMF}), 
which determines the
dependence of the rms radius on the neutron excess parameter $(N-Z)/A$.
The other parameter was held fixed at $\delta=-0.035$ fm.
For antiprotonic atoms 
the two shapes of the neutron distributions lead to almost the same quality
of fit and it is impossible to distinguish between the two on the
basis of the values of $\chi ^2$.
The value of $\gamma$ from pionic atoms is consistent with either values
obtained from antiprotonic atoms but there seems to be a conflict
considering the shape of the neutron density.

To look for the source of this conflict 
it is necessary to look into the radial sensitivities of the two probes,
which have vastly different absorption cross sections in nuclear matter.
This is done with the functional derivative method, introduced originally
in connection with kaonic atoms \cite{BFr07} and later used also for
pionic and antiprotonic atoms \cite{FGa07}. It is shown in \cite{FGa07}
that pionic atom data are sensitive to nuclear densities around the 50\%
region of the central density whereas antiprotonic atom data depend on
the density at the extreme periphery where the densities are well below 10\%
of the central density.

\begin{table}
\caption{Summary of results showing number of
 points and best fit values of $\chi ^2$.}
\label{tab:1}
\begin{tabular}{lllll}
\hline\noalign{\smallskip}
source & shape of $\rho _n$ & N & $\chi ^2$& $\gamma$ (fm)  \\
\noalign{\smallskip}\hline\noalign{\smallskip}
$\bar p$ & `halo' & 90 & 196 & 0.9$\pm$0.1 \\
$\bar p$ & `average' & 90 & 198 & 1.25$\pm$0.15 \\
$\pi ^-$ & `skin' & 100 & 173 & 1.0$\pm$0.1 \\
\noalign{\smallskip}\hline
\end{tabular}
\end{table}

\begin{figure}[t]
\centering
\begin{tabular}{cp{1cm}c}
\begin{overpic}[height=8cm,width=0.47\textwidth,clip]{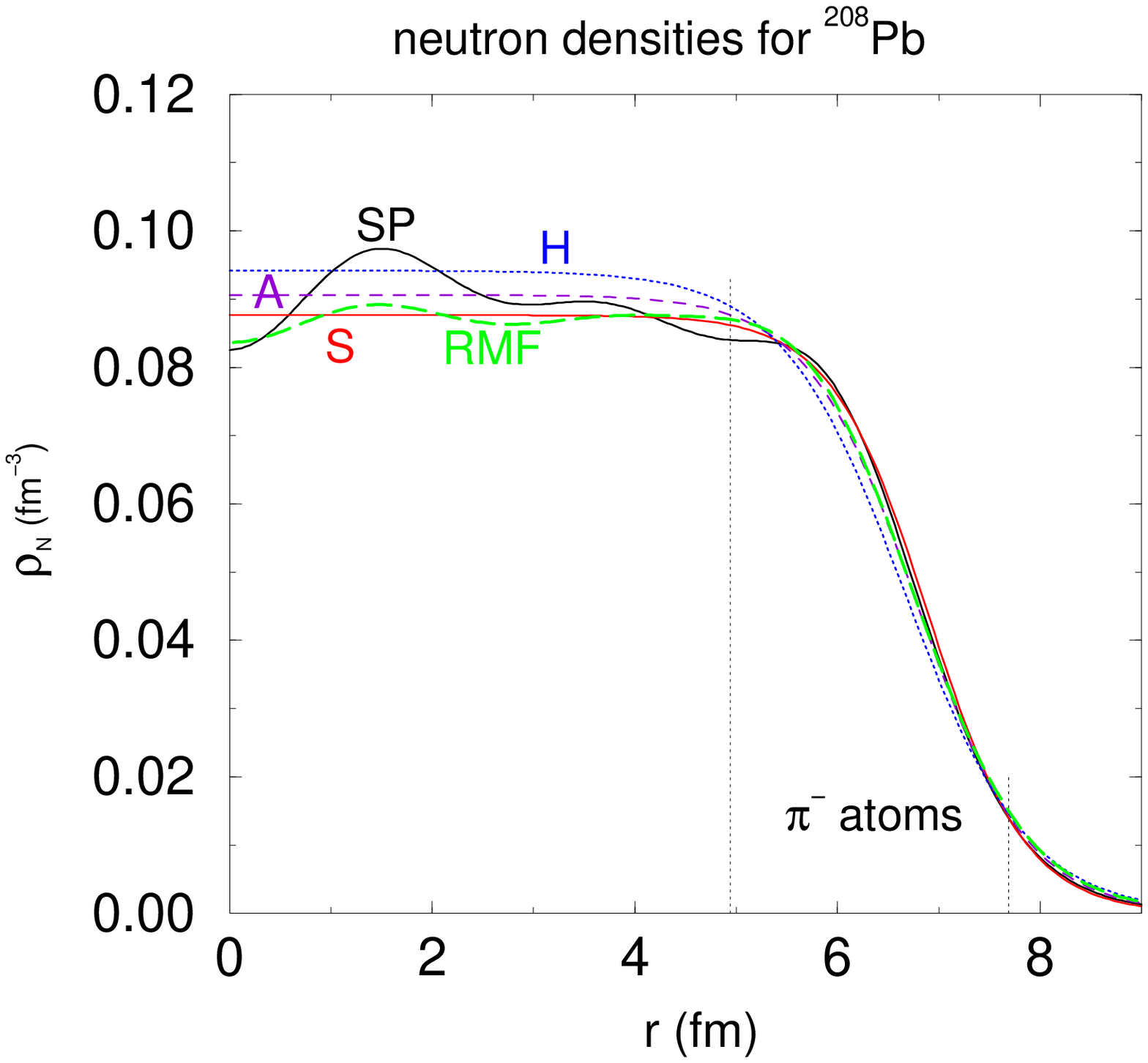}
\end{overpic}
%& &
\begin{overpic}[height=8cm,width=0.47\textwidth,clip]{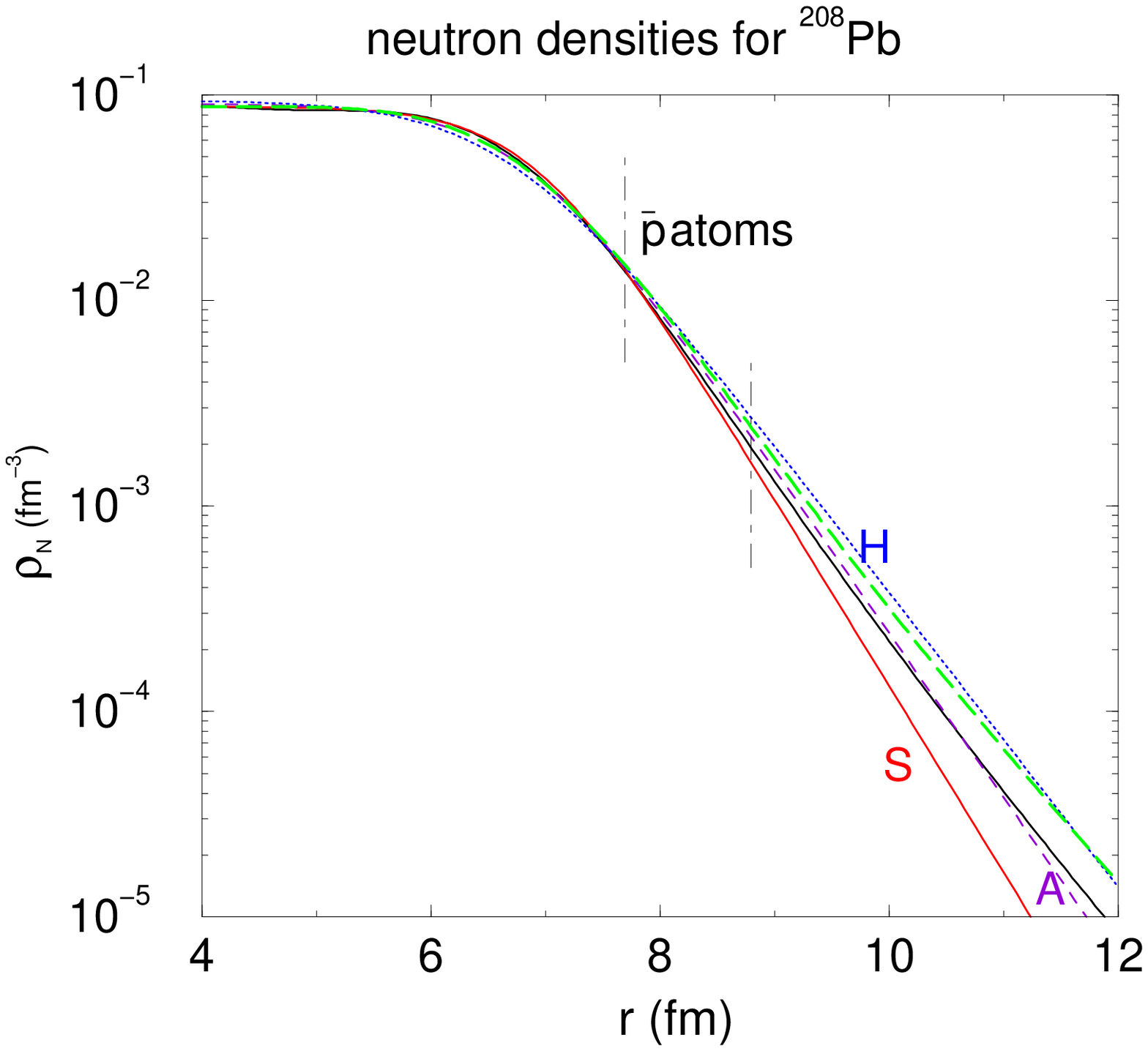}
\end{overpic}
%\\[3.5ex]
%(a) & & (b)
\end{tabular}
\caption{SP and RMF neutron densities in
$^{208}$Pb compared with the three options of 2pF 
parameterizations (S for skin, H for halo and A for average). 
Left: covering the
pionic atoms sensitive region. Right: covering the antiprotonic atoms
sensitive region. All five densities have the same rms radius.}
\label{fig:3}
\end{figure}

Figure \ref{fig:3} shows comparisons between the 
three versions of the 2pF neutron density and two, more physical, neutron 
densities for $^{208}$Pb: 
(i)A single
particle (SP) neutron density obtained by filling in of single
particle levels in a common potential. (ii)A neutron density from a 
more sophisticated RMF calculation \cite{HSe81}.  
All five densities have the same rms radius. On the 
left hand side is indicated the sensitive region for pionic atoms
and on the right hand side is indicated the sensitive region for 
antiprotonic atoms. It is evident that in the pionic atoms region 
the skin and the average shapes
are closest to the SP and the RMF densities whereas in the
antiprotonic atoms region the skin shape deviates markedly from the
two more physical models. At very large radii the halo shape is very close
to the RMF density but at the antiprotonic atoms region 
of sensitivity the average
shape is equally good. 
The conclusion is that the 2pF model is not necessarily able
to reproduce more realistic densities over a broad range of density 
values. In contrast it seems that the values of rms radii are less
sensitive to the model used.
We conclude that there is no real conflict between antiprotonic and
pionic atoms and that on the average the rms radii of neutron density
distributions in nuclei may be represented by Eq.(\ref{eq:RMF})
with $\gamma$=1.0$\pm$0.1 fm and $\delta =-$0.035 fm.
Applying this to $^{208}$Pb which is a most studied nuclide, we find
$r_n-r_p$=0.18$\pm$0.02 fm, in very good agreement with Ref. \cite{KTJ07}.

\begin{acknowledgements}
I wish to thank A.~Gal for many fruitful discussions and J.~Mare\v{s}
for providing some RMF densities.
\end{acknowledgements}

\end{document}